\documentclass[useAMS,usenatbib]{mn2e}
\usepackage[dvips]{graphicx}
\usepackage{amssymb}
\usepackage{txfonts}

\newcommand{\thisgrb}{GRB\,060218}

\newcommand{\chisqred}{\ensuremath{\chi^2_{\mathrm{red}}}}

\newcommand{\alav}{$A_{\lambda} / A_V$}

\title[The extinction properties of GRB host galaxies from H and He\,I ]{The extinction properties of long GRB host galaxies from H and He\,I recombination lines\thanks{Based on observations made with ESO Telescopes at the Paranal Observatory under programme ID 381.D-0723}}
\author[K. Wiersema]{K. Wiersema$^{1}$ \thanks{E-mail:
kw113@star.le.ac.uk}\\
$^{1}$ University of Leicester, University Road, Leicester LE1 7RH\\
}
\begin{document}

\date{Accepted 04 January 2011}

\pagerange{\pageref{firstpage}--\pageref{lastpage}} \pubyear{2010}

\maketitle

\label{firstpage}

\begin{abstract}
In this paper we show how a self-consistent treatment of hydrogen and helium emission line fluxes of
the hosts of long gamma-ray bursts can result in improved understanding of the dust properties in these galaxies.
In particular, we find that even with modest signal to noise spectroscopy we can differentiate different values for $R_V$, the 
ratio of total to selective extinction.  The inclusion of Paschen and Brackett lines, even at low signal to noise, greatly increase the
accuracy of the derived reddening. This method is often associated with strong systematic errors, caused by the need for multiple instruments to cover
the wide wavelength range, the requirement to separate stellar hydrogen absorption from the nebular emission, and because of the dependancy
of the predicted line fluxes on the electron temperature.  We show how these three systematic errors can be negated, by using suitable instrumentation (in particular 
X-shooter on the Very Large Telescope)
and wide wavelength coverage.
We demonstrate this method using an extensive optical and near-infrared spectroscopic campaign of the host galaxy of gamma-ray burst
060218 (SN 2006aj), obtained with FORS1, UVES and ISAAC on the VLT, covering a broad wavelength range with both high and low spectral resolution.
We contrast our findings of this source with X-shooter data of a star forming region in the host of GRB 100316D, and show the improvement over existing
published fluxes of long GRB hosts.
\end{abstract}


\section{Introduction \label{sec:intro}}

As long gamma-ray bursts (hereafter referred to simply as GRBs) are produced by massive stars, as evidenced by their associated supernovae (e.g. Galama et al. 1998; Hjorth et al 2003; Pian et al.~2006; 
Starling et al. 2010; but see also Fynbo et al.~2006), there is 
a clear link between high mass star formation and GRBs. This link is apparent in the properties of their host galaxies: the hosts show bright nebular emission lines associated with active star formation, and show spectral energy distributions and morphologies generally consistent with 
blue, metal-poor, star forming dwarf galaxies (e.g. Savaglio, Glazebrook \& Le Borgne 2009). From high resolution imaging it is evident that the positions of long GRBs in their hosts
are strongly correlated with the location of starformation within these hosts (e.g. Fruchter et al 2006). 
It is therefore a reasonable assumption to make that  the properties of the star forming regions in the host also lead to 
insight in the properties of the progenitor star: the progenitor only lives for a short time. A constraining parameter on the evolution
of the progenitor object is its metallicity,  or more specifically the abundance of iron, as it is the main driver of stellar winds. 
Measuring the gas phase iron abundance from emission lines of galaxies is complicated, as these emission lines are faint and iron is easily and rapidly depleted into dust grains.
Oxygen is a good alternative: its abundance can be easily measured (the lines are bright) either directly (using oxygen lines at various ionisation stages in combination with electron temperature and density sensitive lines); or indirectly through the strong forbidden lines coupled with radiation transport simulations.  
In order to derive accurate abundances, one needs to correct for the extinction experienced by the hot gas that emits the different emission lines. 

Besides deriving element abundances from emission lines, there are further reasons why we are interested in the extinction in the star forming region(s)
in the host galaxies. 
Particularly interesting is the combination of afterglow spectroscopy and photometry, which probes the line of sight dust and gas properties within the host, with host galaxy spectroscopy in emission. 
This combination may constrain the possible destruction of dust by the GRB, it may show how special the region in which the GRB took place is with respect to other star forming regions, and may 
shed light on the (depletion) chemistry in the ISM in GRB host galaxies.
 
In this paper we focus on the extinction in star forming regions within GRB hosts (perhaps better described as the attenuation of emission lines by dust within the hosts). The host galaxy of GRB\,060218 is a particularly attractive target to do this. 
The supernova accompanying this GRB is well studied, and therefore
a connection with host galaxy properties (spectroscopic, spectral energy distribution; see e.g. Savaglio, Glazebrook \& Le Borgne 2009) may provide tight constraints on the progenitor evolution.
In a previous paper (Wiersema et al. 2007; hereafter W07) we performed a study of the metal abundances in this host and the ISM velocity structure, both 
obtained through emission and absorption lines detected in a VLT UVES spectrum of the supernova (2006aj). We obtained three more datasets of this source, which we will use
to probe the extinction properties of this host in this paper. 
These data are part of a dedicated VLT program (program 381.D-0723, PI Wiersema) aimed at a thorough understanding of the stellar population(s) of  
this host galaxy, and consist of optical spectroscopy at medium and high resolution (using the FORS1 and UVES instruments) and low resolution K band spectroscopy (using the ISAAC instrument).

This paper is organized as follows: in Section 2 we describe the way we use H and He\,I line flux ratios to fit the extinction properties; in Section 3 we describe the observations and 
the way we measure emission line fluxes from the spectra; in Section 4 we fit the data using our models and in Section 5 we discuss our results.

Throughout the paper we adopt a cosmology with $H_0 =  71$ km\,s$^{-1}$\,Mpc$^{-1}$, $\Omega_m = 0.27$, $\Omega_\Lambda = 0.73$.

\section{Method \label{sec:method}}
We assume that the foreground absorbing screen of dust is uniform, and express the extinction function in \alav\ and the observed and predicted ratios of two emission lines,
$R_{\rm o}$ and $R_{\rm p}$ at wavelengths $\lambda_1$ and $\lambda_2$: 

\begin{equation}
A_V = \frac{2.5 \log(R_{\rm o} / R_{\rm p})}{(A_{\lambda_2} - A_{\lambda_1}) / A_V }.
\label{eq:ratios}
\end{equation}

We now use the Cardelli, Clayton \& Mathis (1989) parametrization of the extinction curve, 
\[
A_{\lambda} / A_V = a(\lambda) + b(\lambda) / R_V,
\]
with the usual notation $R_V = A_V / E(B-V)$, for a range of $R_V$ values. Note that we will use this on low redshift GRB hosts, so we expect a possible $\sim2200$ \AA\ feature not to have an effect on the analysis.
Any line ratio where the flux ratios can be accurately predicted through atomic physics can be used,  but
it is most common to use hydrogen (H\,I) or helium (He\,I or He\,II) lines because of the large number of available transitions in the optical window, their brightness, the line emissivities can be accurately modelled, and because there is no sensitivity to abundance. 

\begin{figure}   
\includegraphics[width=8.5cm]{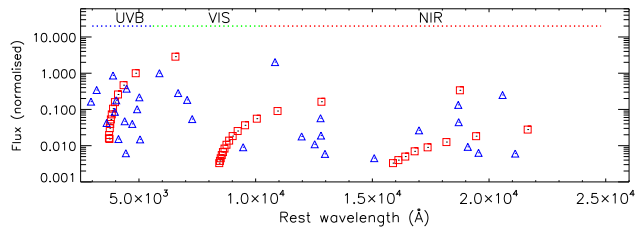}
\caption{This plot shows the position in restframe wavelength of the most useful H and He\,I lines. The hydrogen lines (the hydrogen series is easily recognisable by eye) are indicated by a red square,
with fluxes normalised to H$\beta$ for case B recombination and $T_e = 10^4$ K and $n_e = 100$ cm$^{-3}$. The He\,I lines are shown with blue triangles, their fluxes are normalised to He\,I 
$\lambda$5876, for the same $T_e$ and $n_e$ as  hydrogen. The nominal ranges of the three arms of X-shooter are indicated by horizontal dotted lines. However, we note that the NIR arm efficiency drops very rapidly redward of the K band. It is clear from this plot that the He\,I lines fill in important wavelength gaps left by the hydrogen series -- a majority of the X-shooter spectral orders 
contains one or more emission line.}
\label{fig:hhe}
\end{figure}

He\,II lines are generally too faint in most GRB host star forming regions to be useful in this context, but the He\,I lines do provide meaningful constraints on $A_V$ and $R_V$. In addition, the He\,I  lines fill gaps in the wavelength coverage of the 
H series. As we will demonstrate further below, this is particular useful  for X-shooter spectra: most orders have a H or He emission line in (at least at the low redshifts we consider in this paper).
Figure \ref{fig:hhe} shows this graphically: we plot the wavelength and flux of the relevant H and H\,I lines, where the hydrogen fluxes are normalised to H$\beta$ and the He\,I fluxes to He\,I $\lambda$5876, assuming
case B recombination, $A_V = 0$, $T_e = 10^4$ K and $n_e = 100$ cm$^{-3}$ (see below for details).   
 
We now compute \alav\ for the wavelengths $\lambda$ of all useful H and He\,I lines in the spectra using a range of $R_V$ values from 2.5 to 5.5 (see also Petersen \& Gammelgaard 1997). It is clear that \alav\ changes most strongly with changing $R_V$ for the bluer lines. We then proceed to measure the line fluxes and upper limits on non-detected lines. Hydrogen line fluxes are normalized to H$\beta$, whereas for the He\,I lines we use
He\,I $\lambda$5876: these lines are most frequently detected in low-$z$ hosts and are bright. 
For the predicted flux ratios $R_{\rm p}$ we use a fixed electron density $n_e = 10^2$ cm$^{-3}$: the vast majority of GRB hosts and star forming regions within these hosts show flux ratios of the [O\,II] $\lambda$3727, 3729 doublet
or the [S\,II] $\lambda$ 6716, 6731 doublet in the low density limit. Note also that at low densities the emissivities are fairly weak functions of the density (Osterbrock 1989).
We use a series of different electron temperatures $T_e$, from 0.5 to $2.5 \times 10^4$ K, and obtain the predicted line flux ratios for hydrogen from Hummer \& Storey (1987) for these $T_e$ values and electron density.  
He\,I emissivities are still subject to some debate, but generally the differences in emissivities between different approaches is considerably smaller than the measurement errors we can realistically expect from even the brightest GRB hosts. We use the emissivities and the interpolation over electron temperature as determined by Porter et al.~(2005) and Porter, Ferland \& MacAdam (2007; their appendix A) to compute 
$R_{\rm p}$ for He\,I using the same $T_e$ grid and fixed $n_e$ as for hydrogen.
 
After the fluxes of the hydrogen and He\,I lines in range are measured (detailed in Section \ref{sec:obs}), we arrive for each source at a grid of values of \alav\ (for a range of $R_V$ values) and the corresponding 
$2.5\log(R_{\rm o} / R_{\rm p})$ values (for a range of $T_e$). 
An example for two different values of $R_V$ is shown in Figure \ref{fig:hhefits}. We now fit a straight line to the points, for each combination of $R_V$ and $T_e$,
recording the $\chi^2$. We do this seperately for H and He\,I.  The slope of the best fitting straight line gives the attenuation (it is equal to $-A_V$, see Eq. \ref{eq:ratios}). 
The best description of $R_V$ and $T_e$ is now given by the model with the lowest $\chi^2$.

The use of hydrogen lines to derive $R_V$ and $A_V$ simultaneously is commonly used in galactic emission line regions, see e.g. Greve (2010) and Petersen \& Gammelgaard (1997) and references therein. These authors point out a further simplification that is useful when only limited lines are available: the flux ratios of emission lines from the same species that come from the same upper level are fairly insensitive to $n_e$ and $T_e$.

We now proceed to demonstrate the strengths and weaknesses of the method outlined above for the case of star forming regions in long GRB host galaxies.

\section{Observations}\label{sec:obs}
\subsection{The host of GRB\,060218}
The data of the host galaxy of \thisgrb\  presented in this paper were taken with the aim of characterizing both the young stellar population that produced the GRB itself (e.g. the WR star population) and a possible underlying old stellar population (see W07 for the motivation). In this paper we will restrict ourselves predominantly to the hydrogen and helium lines and the inferred attenuation properties, further physical properties of the source will be discussed in future publications.

General details of our observations are listed in Table \ref{table:logobs}.

\begin{table*}
\caption{VLT observations of the host of GRB\,060218. The seeing is as measured from the acquisition image(s). 
\label{table:logobs}}
\begin{tabular}{lllll} 
Inst. + grism + slit width ('') & Date & Exp time (s) & Seeing ('') & Airmass\\ 
\hline
FORS1 + 600B + 1.3''& 2008-09-04  & $2 \times 1345$                            & 1.5 & 1.3   \\
                                      &  2008-10-01  &  $2 \times 1345 + 1 \times 900$ & 1.1 & 1.3  \\ 
                                      \hline
Inst. + dichroic center wl (filters) + slit width ('') & Date & Exp time (s) & Seeing ('') & Airmass \\ 
\hline                                      
UVES +  437 and 860 nm (HER5+OG590) + 1.2            &   2008-08-02    &  1400                      & 0.9 & 1.4  \\ 
                                      \hline
Inst. + mode + center wl ($\mu$m) + slit width ('') & Date & Exp time (s) & Seeing ('') & Airmass\\ 
\hline                                      
ISAAC + SWS1-LR + 2.2 + 1.0             &  2008-07-17     &   $12 \times 60$              & 0.8  & 1.7  \\
                                                                   &  2008-09-10     &    $108  \times 60$        &  0.4 & 1.5  \\
\hline
\end{tabular}
\end{table*}

\begin{figure*} 
\includegraphics[width=18cm]{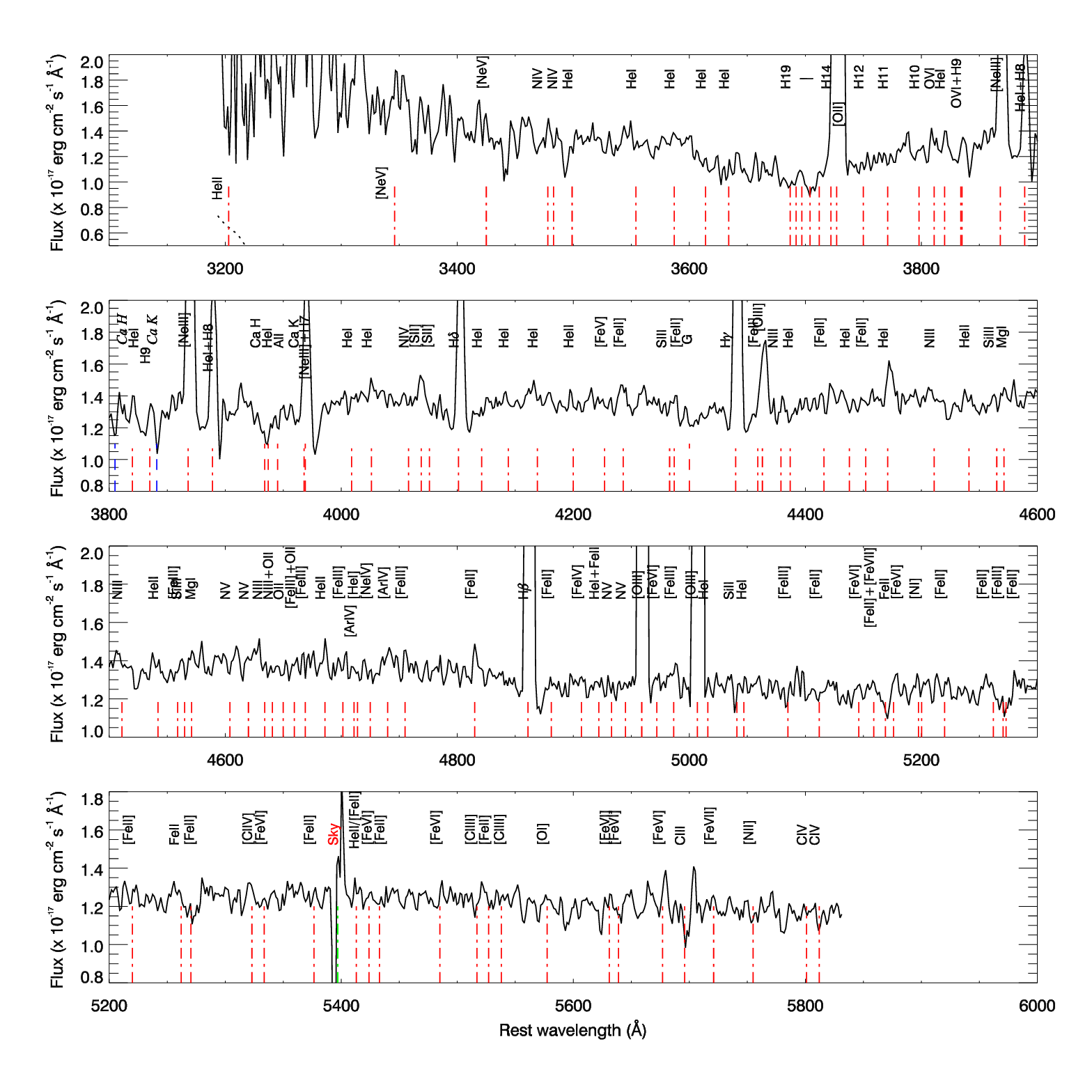}
\caption{The VLT FORS1 600B spectrum (see Section \ref{sec:fors1}) of the host of GRB\,060218, plotted in restframe wavelength scale. 
Indicated by vertical lines are strong and weak (nebular) emission lines, as well as 
some absorption features, commonly detected in star forming galaxies (note that not all the labelled lines are significantly detected). 
Lines labelled in italic font indicate Galactic absorption lines. Clear absorption
components can be seen underneath the hydrogen and helium emission lines. At the scale of this plot, the errorspectrum is not visible (apart from the lowest signal area, 
the dashed line around $\sim$3200 \AA).}
\label{fig:forsspec}
\end{figure*}

\begin{figure*} 
\includegraphics[width=15cm]{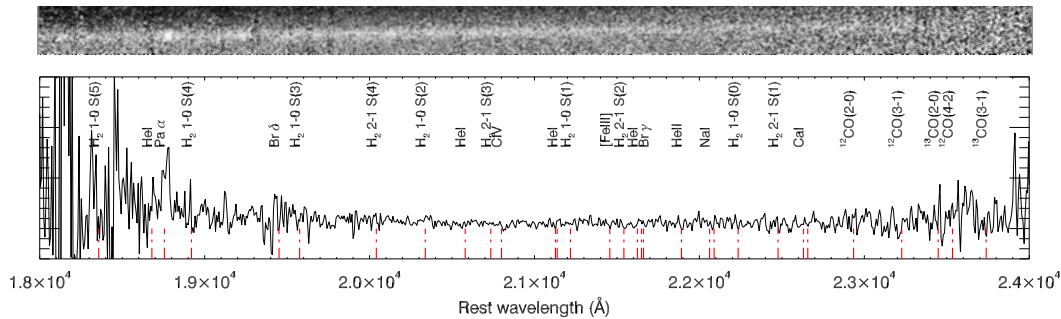}
\caption{The ISAAC spectrum taken with the LR K grism, plotted in restframe wavelengths. The top plot shows the combined 2 dimensional spectrum before 
telluric line correction and sensitivity calibration. The lower graph shows the extracted, telluric line corrected and flux calibrated spectrum. 
Several emission and absorption lines that are commonly detected in (ultracompact) HII regions, starburst galaxies and WR stars are indicated. 
Whereas continuum emission is detected over nearly the entire spectral range, most of the lines indicated are not significantly detected. 
}
\label{fig:isaacspec}
\end{figure*}

\begin{figure}   
\includegraphics[width=8cm]{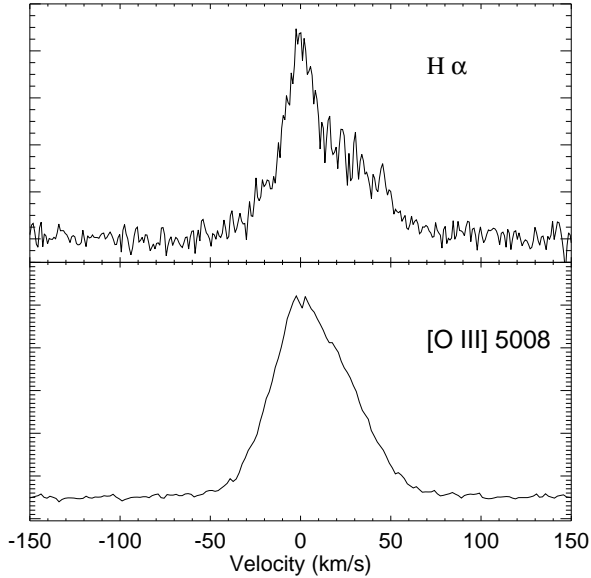}
\caption{The top panel shows the H$\alpha$ emission line from the UVES data described in Section \ref{sec:uves}; the lower panel shows the [O\,III] $\lambda$5008 emission line from W07.
In addition to these lines, several other emission lines  are resolved.}
\label{fig:resolved}
\end{figure}

\subsubsection{FORS1 spectroscopy \label{sec:fors1}}
We targeted this host with FORS1, as this instrument has been blue optimised (the similar FORS2 instrument is red optimised). The 600B grism was used. While this grism 
does not give us coverage at the red end of the visual range, its higher resolution compared to e.g. the 300V grism (a study of this host with this grism is presented in Han et al.~2010)
allows more accurate deblending of blue Balmer lines and more accurate fits for stellar absorption. In addition, the higher resolution, though still low, allows a more accurate view of the 
emission lines making up the WR star bump claimed by Han et al.~(2010). 
We selected a 1.3 arcsecond slit. The width of the slit was chosen based on high resolution imaging from the Hubble Space Telescope (HST), see e.g. Starling et al.~(2010) for an HST image of the host, so 
that we are more likely to get all the galaxy light in the slit.

Observations were performed at two epochs, with poor seeing conditions in the first epoch. Acquisition was done using direct on-source acquisition on the source photo-center, as the host is bright
for a GRB host ($r = 20.16 \pm 0.03$; Sollerman et al.~2006), the star forming region(s) (see Section \ref{sec:uves} and W07) are located at or near the center of the host galaxy.

The data were reduced using standard routines in {\sc IRAF}, using routines in the {\it ccdproc} and {\it specred} packages, using calibration data (bias frames, flatfields, arc lamp frames and standard star exposures) taken the same nights as the science data. 
Flux calibration of the data of Sep 4 was achieved using exposures of F type spectrophotometric standard star LTT 1788 (Hamuy et al. 1992, 1994). The standard star for the second night of data
was DA4 star HZ 4 (Oke 1990).  The extracted, flux calibrated, individual spectra from the two nights were combined through a weighted average, using the signal to noise per bin as weight. As expected, the signal to noise is dominated by the second night spectra, that have considerably better seeing conditions.  We applied a mean extinction curve for Paranal, and dereddened the spectra for the Galactic extinction value of $E(B-V)_{\rm Gal} = 0.142$ and $R_V = 3.1$, see for a discussion W07.
Published photometry of the host (Sollerman et al.~2006) was used to bring the spectrum to an absolute scale. We measure a continuum signal to noise of $\sim50$ per pixel 
and a arc line FWHM (as a measure for the resolution) of 5.0 \AA\  (at $\sim$5000 \AA).

A multitude of emission lines are detected, see Figure \ref{fig:forsspec}, where we have labelled lines commonly found in high signal to noise spectra of blue compact galaxies and WR galaxies (many labelled lines are not detected). In the following we will only discuss the hydrogen and He\,I lines, the metal lines will be discussed in a future publication (Wiersema et al in prep.).

\subsubsection{ISAAC spectroscopy \label{sec:isaac}}
We acquired deep low resolution ISAAC K band spectroscopy, see Table \ref{table:logobs}.
 Data were reduced using the ESO ISAAC data reduction suite, version 5.7.0, using calibration data (dark frames, flatfields, arc lamp frames) taken the same night. Reduction included sky  subtraction, flat fielding, wavelength calibration, spectral image  registration and co-addition.
The data were taken using 30 arcsecond nod throws, and a 10 arcsecond jitter box width. Wavelength calibration from the arc line frames was refined using
night sky lines, and the host galaxy spectrum was extracted using tasks in {\sc IRAF}.  Observations of the B3V star Hip 021575 (HD 29376) were taken for telluric line calibration:
telluric correction and flux calibration was achieved using the IDL / SpeXtool package (Cushing, Vacca \& Rayner 2004), in particular the {\em xtellcor\_general} task, which uses  a high quality Vega model empirically convolved with known telluric absorption features to match the entire telluric standard spectrum in detector units (thereby implicitly solving for the instrument response). The resulting model is then applied to the science target spectrum, producing a  telluric feature corrected and flux calibrated spectrum (excepting slit losses). To bring the spectrum onto an absolute flux scale, one generally uses 
photometry of the source. In the case of this source, Kocevski et al.~(2007) give $K_s = 18.73 \pm 0.34$, uncorrected for Galactic extinction. However, the Kocevski et al.~measurements were done with poor spatial resolution. We therefore examine the reduced acquisition images from the second epoch of ISAAC data, which provide excellent pixel scale under good seeing conditions.
We combine the three 10 second acquisition exposures ($K_s$ filter), and tie the astrometry to the FORS1 acquisition images. The host galaxy is detected, and is resolved, but detection
significance is low. We perform aperture photometry  with respect to four 2MASS stars in the field. We find $K_s = 18.3 \pm 0.3$ for the host, but because of the poor detection rate and the small number of 2MASS stars in the field, we use the Kocevski et al.~value to bring the spectrum to an absolute scale. 

The slit position angle was chosen such that a nearby 2MASS star fell into the slit, with the aim of providing a more accurate flux calibration in case the acquisition data were not deep enough to detect the host, particularly since the value from Kocevski et al.~(2007) has a large uncertainty. However, the spectrum and acquisition images show that this object, 2MASS 03213974+1651305 ($K = 12.55$) is in fact
a blend of two stars, separated by 0.9 arcseconds, and with the components not fully covered by the slit. We are therefore unable to further finetune the absolute flux offset of our spectrum using this object.

\subsubsection{UVES spectroscopy \label{sec:uves}}
We acquired a high resolution spectrum of this source with the aim of verifying the multiple velocity components observed in both emission {\em and} absorption ([O\,III] and Na\,I + Ca\,II, respectively) in 
UVES spectroscopy of SN\,2006aj, the supernova accompanying GRB\,060218 (W07). 
We selected the central wavelengths of the spectra such that we would cover both the H$\alpha$ emission line and the [O\,II] doublet in the red and blue arms, lines that will aid in detecting
changing physical parameters over the various velocity components (see also Section \ref{sec:caveats}). 

Because of a spatial coordinate offset error only the exposures taken on Aug 2 can be used here.  We reduce the data using the standard prescriptions within the ESO UVES data reduction package, version 4.2.4. A crude flux calibration is achieved using spectrophotometric standard stars (EG 274; Hamuy et al. 1992, 1994) taken on Aug 5 and Aug 8, no spectrophotometric standards were observed on the night the data were taken. Because of the short exposure time, no continuum emission is detected, but we do detect a dozen emission lines, including the three brightest Balmer lines and the [O\,II] lines. In this paper we only use the velocity information in some of the lines, and further calibration of the line fluxes through the FORS1 data and the data in W07 will be done in future papers.
The UVES spectrum is converted into vacuum wavelengths using the IRAF {\em disptrans} task and brought onto a heliocentric wavelength scale, after which we transform the wavelengths onto the
solution of W07 by measuring the wavelength offsets of six lines detected in both the current UVES data and the UVES data in W07. This allows us to directly compare the 
profiles of the detected emission lines with the ones in the W07 UVES data on a fixed relative velocity scale. In Figure \ref{fig:resolved} we show a comparison of the H$\alpha$ emission line
from the data described above with the [O\,III] $\lambda$5008 line as analysed in W07. The complex line morphology seen in both emission and absorption is likely caused by two or more 
star forming regions and/or a bright star forming region with outflow(s) and galaxy disk-like emission. 
The HST imaging data are not able to resolve this, but the detection of several resolved emission lines may enable us to constrain these models somewhat (Wiersema et al in prep.).

\begin{figure}   
\includegraphics[width=8cm]{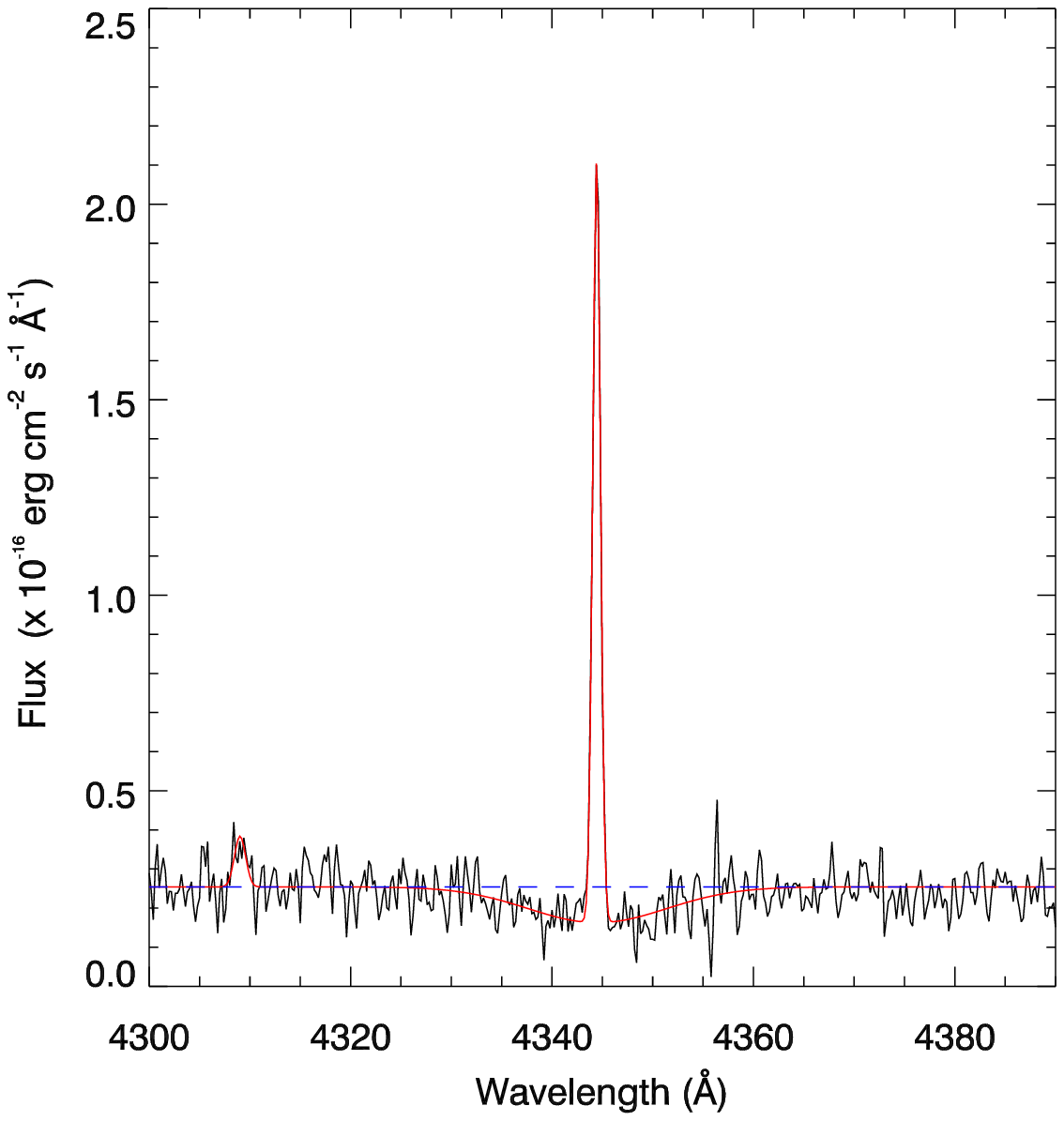}
\includegraphics[width=8cm]{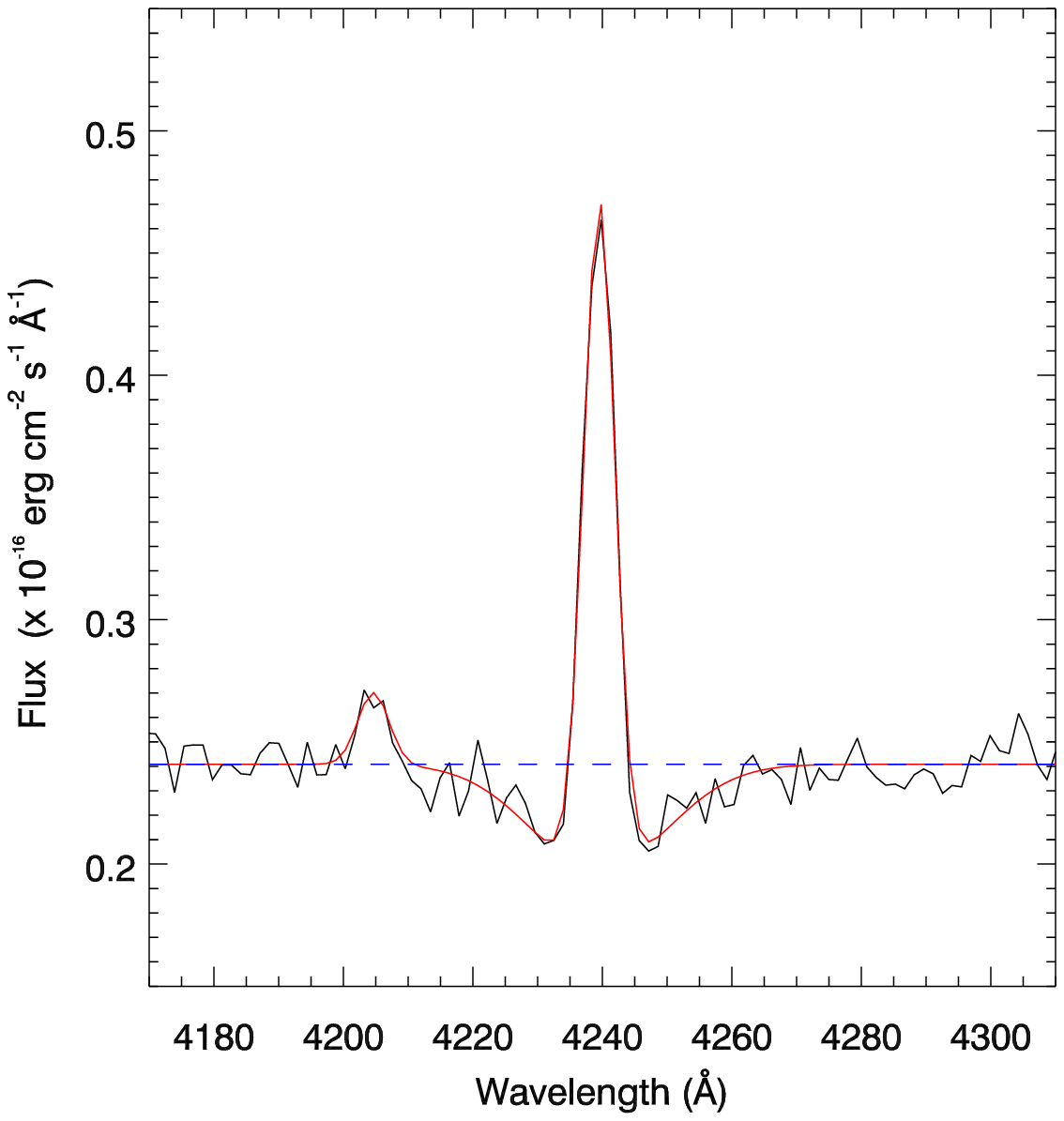}
\caption{Fit of the H$\delta$ lines of the X-shooter spectrum of source A in the host of GRB\,100316D (top) and the FORS spectrum of the host of GRB\,060218 (bottom). The weak emission line blueward
of H$\delta$ in both spectra is [S\,II], which is included in the fit. 
The line is clearly made up of an emission component (formed by hot gas) and an absorption 
component (from stellar atmosphere absorption). Both emission and absorption lines are adequately fit by Gaussian profiles. The resulting fit is shown in red, the blue dashed line indicates the 
continuum level.  The H$\delta$ emission line fluxes for these two sources are virtually identical, while the absorption equivalent width of the host of GRB\,060218 is significantly lower.
This plot demonstrates the necessity of adequate spectral resolution.}
\label{fig:hdelta}
\end{figure}

\subsection{The case for X-shooter: source A in the host of GRB\,100316D}
To demonstrate the advantages X-shooter offers compared to our observations with multiple instruments of the host of GRB\,060218, we make use of the spectra we obtained of a star forming region in 
the host galaxy of GRB\,100316D. This star forming region is not the one hosting the GRB, but simply serves as an illustration of the methods used. The observations, the data reduction and calibration
and the basic properties of this region are reported in Starling et al.~(2010). 
More detailed properties of this host, including the star forming region where the GRB occured, will be reported in a future paper (Flores et al. in prep.).

\subsection{Line flux measurements \label{sec:linemeasure}}
Besides the narrow nebular emission lines, the spectra also show absorption components.
As demonstrated for H$\delta$ in Figure \ref{fig:hdelta}, a clear absorption component is present underneath the emission line, made by stellar atmosphere absorption. The strength (equivalent width) of the absorption component of the line is a strong function of the age of the stellar population dominating the observed luminosity, the starformation history (starburst, continuous) and the metallicity (e.g. Gonzalez Delgado \& Leitherer 1999). Most of these properties are poorly constrained or completely unknown for GRB hosts.
Subtracting a constant equivalent width of all integrated hydrogen line fluxes to account for absorption (as done in Levesque et al. 2010 and for some sources in Savaglio, Glazebrook \& Le Borgne 2009) will lead to inaccurate line fluxes and create inconsistencies when inferring stellar population properties. More accurate is the method employed by Han et al.~(2010), who fit a linear combination of single stellar population models from the evolutionary population synthesis code of Bruzual \& Charlot (2003). As GRB star forming regions are often considered bursty, metal poor and the dominant stellar population is likely very young (as the GRBs form from the massive end of the IMF), we prefer not to rely on models. We perform free fits on the continuum, emission and absorption lines, accepting this may come with larger errors on the emission line fluxes. The choice of spectral resolution and required signal to noise for the 060218 FORS1 data was motivated by the necessity of fitting absorption line components in this manner.  

The emission and absorption components are adequately fit by Gaussian profiles (except in the UVES data of the host of GRB\,060218, see Section \ref{sec:caveats}). We fit each line region separately   
using the {\em ngaussfit} task in the {\em stsdas} package within {\sc IRAF}. We started the fitting with the H$\delta$ line, whose absorption and emission components are easy to separate and have good signal to noise. As demonstrated in Figure \ref{fig:hdelta} we fit simultaneously the continuum (as a first order polynomial); the emission line center wavelength, Gaussian full width at half maximum (FWHM) and amplitude; and absorption line  center wavelength, Gaussian FWHM and amplitude. 
Errors on these fitted parameters are calculated by resampling, which we use as checks on fit quality. From these parameters we compute the emission and absorption line fluxes and equivalent widths, and their $1\sigma$ uncertainties. As can be seen in Figure \ref{fig:hdelta}, the FWHM of the absorption line is considerably larger than that of the emission line, allowing in most cases reasonable fits.
From these free fits on the H$\delta$ lines we found the absorption and emission line to be centered at the same wavelength to within errors. In the fits of lines with relatively poor signal to noise we therefore tie the emission and absorption line centers together in the fits. In several cases, most notably the Balmer lines blueward of H$\delta$, the H or He\,I emission lines are blended with other emission lines, or the underlying hydrogen and helium absorption lines are blended with other, unrelated, absorption lines. In several cases it was possible to deblend by fitting an additional Gaussian fixed at the wavelength of the 
additional component (for example the H$\epsilon$ transition is blended with [Ne\,III]), but in some cases the resolution of the FORS1 data was not sufficient to obtain accurate emission and absorption fluxes, and we do not use these lines in further analysis. We will report the absorption equivalent widths and their implications for the stellar populations in a future paper.

In the near-infrared we further exclude lines that are affected by sharp residuals from telluric line corrections, or, in the case of source A in the host of GRB\,100316D, by bad pixels. We further exclude 
lines from the analysis that have very large errors (in some cases this is because of proximity to the edges of the echelle orders where signal to noise occasionally drops steeply, or when the 
line is right in an area where telluric absorption brings the response to near zero).

\section{Fits of the extinction \label{sec:analysis}}
\subsection{Literature Balmer flux data \label{sec:hanetalfits}}
\begin{figure}   
\includegraphics[width=6.5cm]{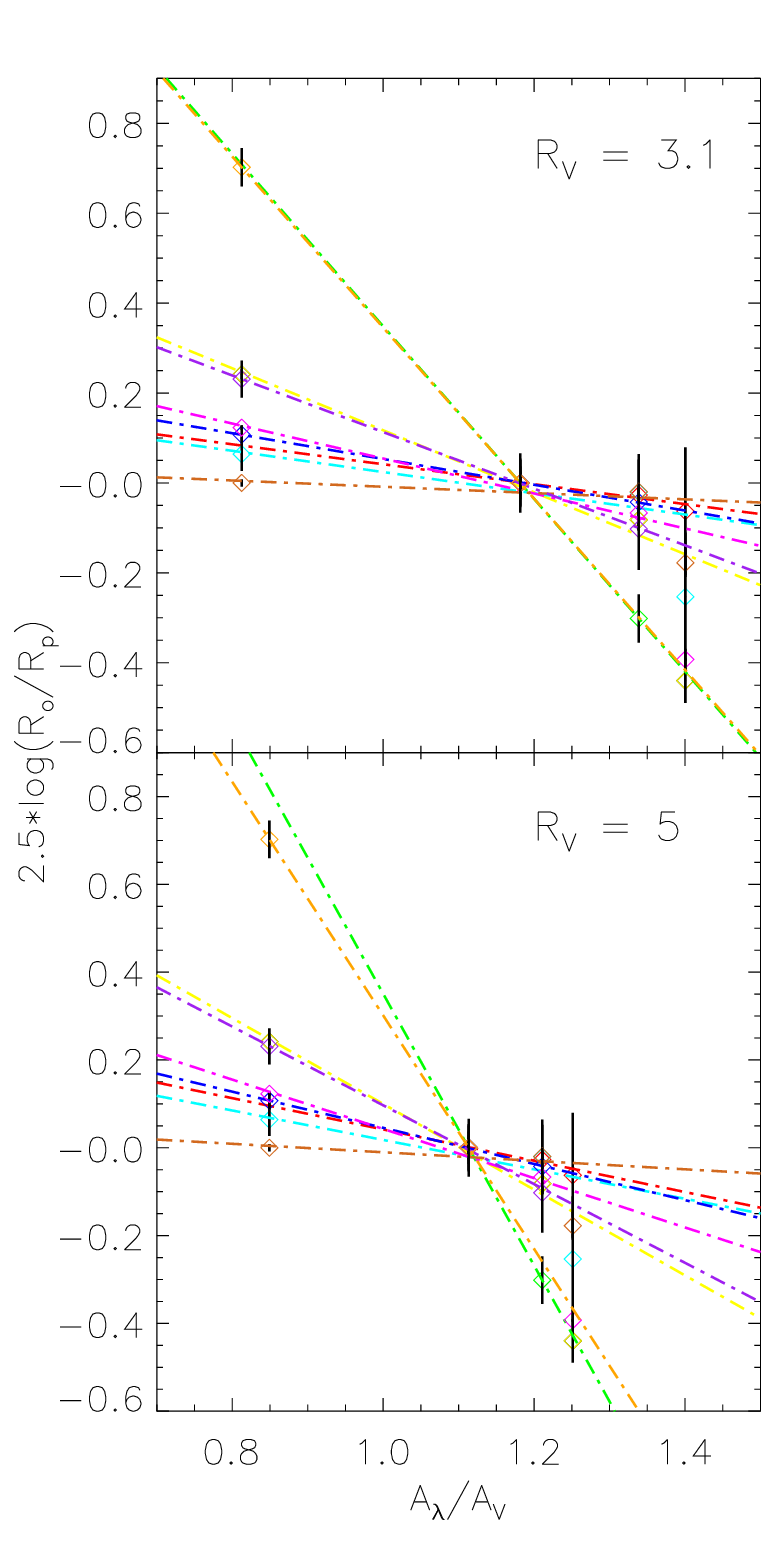}
\caption{An example of one of the fits on the Balmer H$\alpha$, H$\beta$, H$\gamma$ and H$\delta$ emission lines from a series of long GRB hosts from Han et al (2010). 
Both plots show a fit using $T_e = 10^4$ K, with in the top plot $R_V = 3.1$, and the lower plot $R_V =  5$. The $A_V$ values that come from these fits are shown in the panels. These fits are not necessarily the ones with the best $\chi^2$, but nicely demonstrate the influence of
$R_V$ on the fit.}
\label{fig:hanetalsample}
\end{figure}

Several optical spectroscopic studies have been performed on the nearby (long) GRB host galaxy population, generally with the aim of comparing the GRB host population
with other samples of galaxies, see for example Han et al. (2010), Levesque et al. (2010). Han et al fit stellar population models, thereby accounting for stellar absorption, and list the fluxes of the H$\alpha, \beta, \gamma$ and $\delta$ lines for eight host galaxies (for the host of GRB\,060505 both the values for the integrated host spectrum and the region where the GRB took place are given;
Th\"one et al. 2008).
Han et al.~(2010) compute the extinction by assuming $R_V = 3.1$ and $T_e = 10^4$ K, using the ratios of two lines at a time (e.g. H$\alpha$ / H$\beta$). 
Figure \ref{fig:hanetalsample} shows our fits on these emission line fluxes for two values of $R_V$ and $T_e = 10^4$ K. As in our method the fluxes are normalised to H$\beta$,
the fits to all sources go roughly through the same point, and the slope is equal to $-A_V$. It is clear from this figure that most GRB hosts in the Han et al.~sample have low $A_V$. This can not necessarily be extrapolated to all GRB hosts in general: there is a bias against dusty sight lines in the detection of afterglows (see e.g. Greiner et al. 2011), which may in turn translate in a bias against dust rich host galaxies.
 
Of the sources in this sample, only four have four Balmer line detections (the hosts of GRBs 990712, 020903, 031203, 060218), making it impossible in most 
cases to distinguish the effects of $R_V$ and $T_e$ through fit $\chi^2$ (we have two free parameters to fit for each point in our $T_e$, $R_V$ grid: the slope and offset of the
straight line). Of the sources with four detected lines, all but 020903 have poor fit $\chi^2$, in most cases dominated by a single emission line (often H$\delta$).

The fits for 020903 have good reduced $\chi^2$ values, but no preference for a $R_V$ or $T_e$ is clear from this data (the reduced $\chi^2$ ranges from 0.93 to 1.05) because of
the low $A_V$ (in the range $A_V = 0.23$ to 0.33 for the whole $R_V$, $T_e$ grid) and the small \alav\ range. It is clear from these fits that we require
significantly more lines and \alav\ range to constrain these parameters.
In addition we note that the Han et al. sample deals mainly with host-integrated spectra (with the exception of GRB\,060505), and since each host may contain multiple bright
star forming regions, the measured attenuation is a superposition of these, and not easily coupled to GRB environment parameters (i.e. the influence of WR stars associated with the starformation episode
producing the GRB progenitor). As we will show in Section 5, this may also
be evident in our data on the host of GRB\,060218 through our high resolution spectroscopy.

\begin{figure*}   
\includegraphics[width=16cm]{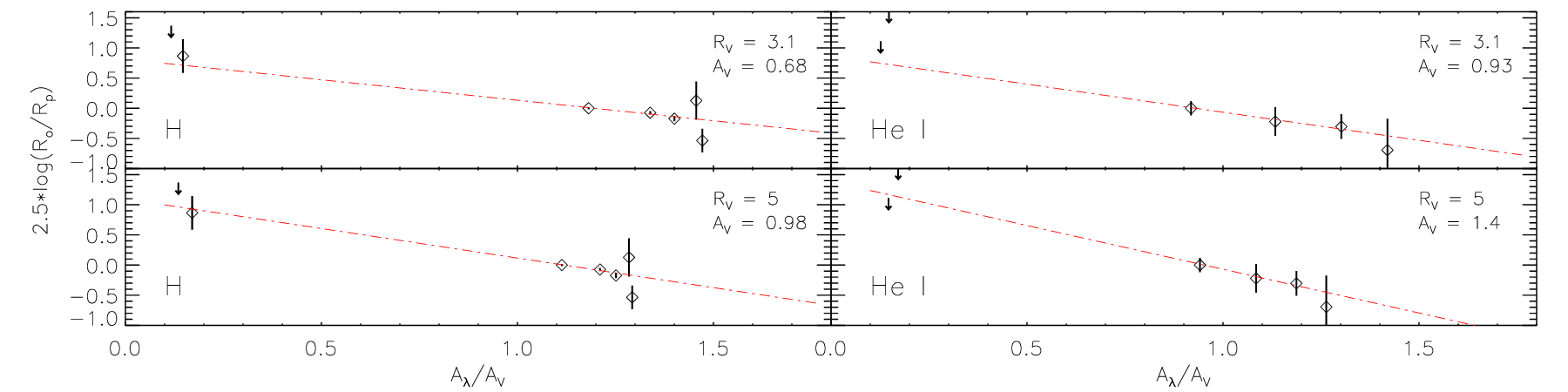}
\includegraphics[width=16cm]{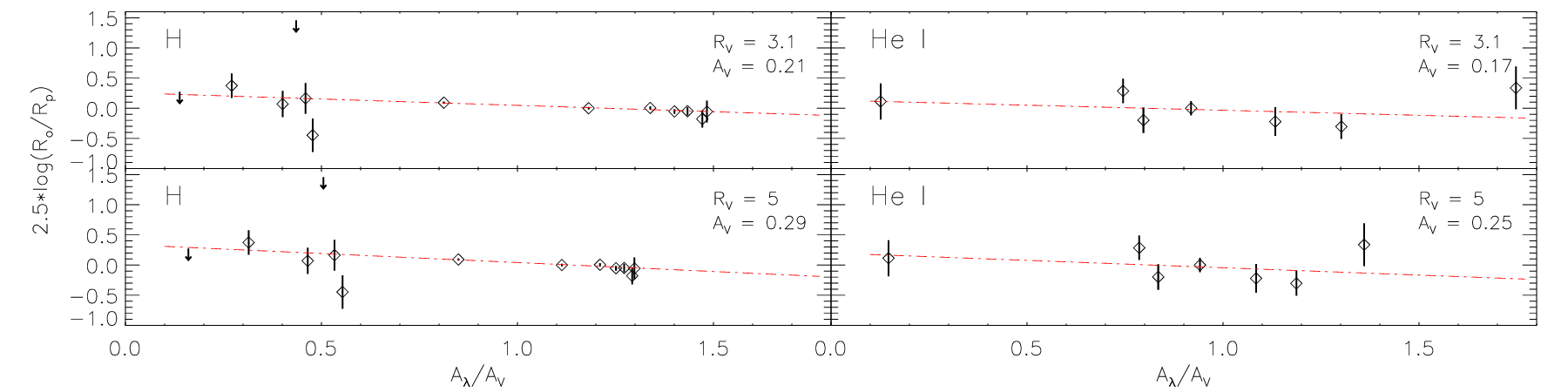}
\caption{
An example of two of the fits in the $T_e, R_V$ grid on the data of the host of GRB\,060218 (top four panels) and source A in the host of GRB\,100316D (lower four panels). 
The left column shows a fit using the hydrogen lines, with $T_e = 10^4$ K, displayed are the fits for $R_V = 3.1$ and 5. The right hand column shows a fit of the 
neutral helium lines, using the same parameters. Indicated are the corresponding best fit values for $A_V$. These fits are not the ones with the best $\chi^2$ for either of 
the sources, but serve to demonstrate the influence of $R_V$ on the fit, and $T_e = 10^4$ K allows easy comparison with literature values. Note that the fits displayed for 100316D 
use also the bluest He\,I line in the fit.}
\label{fig:hhefits}
\end{figure*}

\subsection{The host of GRB\,060218 \label{sec:060218fits}}
It is clear from the fits in section \ref{sec:hanetalfits} that a larger \alav\ baseline is beneficial. The data of the host of GRB\,060218 are well suited to demonstrate 
the possibilities. This host has been observed with HST (e.g. Starling et al.~2010) showing it to be be a blue compact galaxy which appears to contain one bright massive star formation region
(but see Section \ref{sec:caveats} and Wiersema et al in prep.).  

Figure \ref{fig:hhefits} (top) shows an example of two of the fits, at $R_V = 3.1$ and 5.0, and $T_e = 10^4$ K, for the hydrogen and helium lines detected in the FORS1 and 
ISAAC data described above. 
Note that because of the relatively low resolution of the FORS1 600B grism we are not able to derive accurate fluxes of all detected hydrogen lines
because of blending with other emission lines (e.g. H$\epsilon$) and/or absorption lines (lines blueward of H9), this is easily visible in Figure \ref{fig:forsspec}. 
This is important: even though the lines blueward of H$\delta$ get ever closer together in \alav\ space (see Figure \ref{fig:hhe}), this part of the \alav\ range is most sensitive to 
$R_V$ changes. A further caveat is the flux of the Pa$\alpha$ line, whose uncertainty is large because of the large uncertainty in the host $K$ band magnitude. 

Using the fluxes as determined above we find acceptable fits for the hydrogen lines for most models on our $T_e, R_V$ grid, with the lowest $\chi^2$ for the $T_e = 1.5 \times 10^4$ K
and $R_V = 4.5$ model (\chisqred $= 1.3$). The $\chi^2$ distribution shows that the models with high $R_V$ values perform better than models with low $R_V$ ($\lesssim3.5$) for all
temperatures considered. For all models with $R_V \gtrsim 3.1$, higher temperatures ($T_e > 12500$ K) fit better than lower ones. 
Broadly speaking these results indicate a preference for a high temperature, high $R_V$ attenuation, consistent with the findings in W07 and a ``grey'' extinction curve.
The best fitting model corresponds to an attenuation $A_V = 1.02 \pm 0.16$ magnitudes. 
We fit the He\,I lines seperately and find, as expected from the small number of lines detected and the large error bars, that we can not reliably distinguish $R_V$ and $T_e$ preferences:
reduced $\chi^2$ values range from 0.2 to 0.4. We do see
that the $A_V$ values found from the He\,I line fits are consistent with the hydrogen ones. Adopting the same parameters as for the best hydrogen fit we get $A_V = 1.3 \pm 0.7$. 

While these fits show the possibility to distinguish models with different $R_V, T_e$ from a relatively small number of lines, we note that the difference in $\chi^2$ between the best and worst fitting model ($R_V=2.5, T_e = 2 \times 10^4$) is only 1.7. A larger number of detected lines, particularly around \alav $\sim 0.5$ and blueward of H$\delta$, can improve this considerably. 
Taking the best fitting model at face value, the preference for a $R_V$ larger than the mean Galactic value of 3.1 is interesting. Large values of $R_V$ are indicative of sightlines dominated by relatively large dust grains. Dust is extremely sensitive to environmental properties, as evidenced by the wide variety of Galactic extinction law shapes (Valencic, Clayton \& Gordon 2004), and fairly grey extinction curves may point to regions with massive stars (and/or dense environments with shock effects) and/or fairly incomplete dust mixing and processing through the galaxy.

 We caution that
the UVES spectroscopy gives some tentative hints for more than one star forming region in this host (perhaps two, if the Gaussian components making up the line are two distinct star forming regions), which was not evident in HST imaging. 
This may imply that the fits we perform here are of the superposition of the dust properties of these two regions, with potentially different $R_V$ and $A_V$. We therefore will not compare the dust properties
found above further with those found from absorption spectroscopy.

\subsection{X-shooter data \label{sec:xshooterfits}}
HST imaging of the host galaxy of GRB\,100316D shows that the source we refer to as source A (see Starling et al.~2010 for details of this source) is a single, large star forming region unrelated to the
GRB. We use the data of this source only to show the possibilities X-shooter offers in the study of dust in GRB hosts. 

We fit the fluxes determined in Section \ref{sec:linemeasure}. In this case, we do detect He\,I lines at red wavelengths, and because of the higher resolution of X-shooter compared to FORS1 600B grism,
we can de-blend several of the bluer Balmer lines (but not H8). Unfortunately, the redshift of this host galaxy ($z = 0.0591$, Starling et al.~2010) is somewhat unfortunate: several of the brighter lines in the near-infrared fall 
on areas heavily affected by telluric absorption (e.g. Pa$\gamma$), on bad pixels (e.g. Br$\epsilon$) or are redshifted out of the window (wavelengths redwards of $\sim2.26$ $\mu$m are strongly affected by thermal noise and left out of the analysis, this affects e.g. Br$\gamma$). We further note that no infrared imaging of source A was available at the time of writing to assess the quality of the flux calibration. 

The hydrogen lines are well fit by the models, see Figure \ref{fig:hhefits} for two examples which clearly show a significantly lower attenuation than in the host of GRB\,060218. 
This low attenuation means there is little sensitivity to $R_V$ and $T_e$. The $\chi^2$ has a broad minimum at \chisqred$ = 0.86$ for $T_e = 7500$ K and $R_V = 3.5$. For all values of $R_V$ the $\chi^2$ increases with increasing electron temperature (to a maximum of 8.7 at 20000 K, i.e. 1.0 above the best fit), allowing us to conclude the electron temperature in source A is likely $\lesssim 10^4$ K. For each temperature there is very little change in $\chi^2$ with changing $R_V$, the entire range $R_V = 2.5 - 5.5$ only changes $\chi^2$ by 0.1, which is expected when $A_V$ is very low. 
The $A_V$ found for what is formally the best fitting model ($T_e = 7500$ K, $R_V = 4.5$) is $A_V = 0.19 \pm 0.06$. 
 
The He\,I fluxes are also well fit with the models (see Figure \ref{fig:hhefits} for an example). We fit the data set twice, including and excluding the He\,I line at restframe 3187 \AA: this line is observed at 3376 \AA\ and the validity of the mean Paranal atmospheric extinction curve is uncertain at these very blue wavelengths. The fact that this line is even detected is quite remarkable, and will prove valuable for GRB hosts at higher redshifts, as it extends sensitivity beyond the Balmer limit (see Figure \ref{fig:hhe}). 
We find that including this uncertain line flux in the fit increases the $\chi^2$ significantly (doubling it for most models), but we show a fit including this line in Figure \ref{fig:hhefits} for completeness.  
Excluding this line, the He\,I line fluxes show broad agreement with the H results, but similar to the H lines the $\chi^2$ does not change much: for the $T_e = 10^4$ K models the $\chi^2$ changes from 
3.2 to 3.9 from $R_V = 2.5$ to 5.5 . The best fitting model ($T_e = 10^4$ K and $R_V = 3.0$) gives \chisqred $= 0.8$ for $A_V = 0.43 \pm 0.31$.

\section{Discussion} 
\subsection{Caveats, implicit assumptions and calibration requirements \label{sec:caveats}}
In the list below we briefly outline a few important caveats, implicit assumptions and calibration requirements to allow the fits as described above to be successful. We illustrate these
through the fit results of the two sources above.
\begin{itemize}
\item {\bf Signal to noise}: As can be seen from the poor results described in Section \ref{sec:hanetalfits}, we require good signal to noise to detect also the weaker blue Balmer lines
and the weaker Paschen and Brackett lines. The helium lines are even fainter than the hydrogen lines, but form an important addition particularly when infrared spectroscopy is not possible
or when the redshift is fairly high (Figure \ref{fig:hhe}).
\item {\bf Redshift}: the signal to noise requirements naturally translate in a redshift requirement, but the redshift plays a more important role in the position of the emission lines. The fits above show the
necessity of detection of one or more infrared lines (e.g. Pa$\alpha - \delta$ and Br$\gamma, \delta$), which become increasingly hard to detect with increasing redshift.
\item {\bf High resolution imaging}: the redshift plays a role in this requirement too. We require high resolution (i.e. space based imaging) to establish exactly which regions of the host are covered by the slit,
and to make sure we are not averaging multiple star forming regions together. Additionally, a comparison with dust properties in absorption from the afterglow spectra will require understanding of the
spatial position of the GRB within the host. 
\item {\bf Spectral resolution}: As discussed in Section \ref{sec:linemeasure} and Figure \ref{fig:hdelta}, a high spectral resolution and signal to noise is required to effectively fit the absorption components
under the H and He\,I lines, and to deblend these lines from other emission and absorption lines, This is particularly important for the higher Balmer lines, which are vital to diagnose $R_V$.
There is an additional use of high resolution spectra, as demonstrated in Figure \ref{fig:resolved}: the resolved [O\,III] lines (confirmed by the H$\alpha$ data) can be explained by two star forming regions (W07). The detection of two velocity components in absorption in the supernova spectra indicates that in this scenario the GRB is located in the redmost component. However, there are alternative scenarios to explain the multiple components (Wiersema et al in prep). In the case of our analysis of the host of GRB\,060218 the UVES spectra may imply that we are really seeing the
average properties of two star forming regions in the analysis in Section \ref{sec:060218fits}.

\item {\bf Calibration}: The wide wavelength range required for this analysis often means acquiring data with more than one instrument (but see \ref{sec:Xshooteradv}). This puts requirements on the
absolute flux calibration: within one instrument (or X-shooter arm) a relative flux calibration will be sufficient as we consider flux ratios, but when considering multiple instruments, an absolute calibration
is required. Each wavelength range offers its own problems, in addition to the common problems with absolute flux calibration (spectrophotometry), illustrated above. Particularly in the infrared this can be very difficult, as the telluric transmission has to be fit for in addition to the sensitivity, slitlosses and extinction, putting additional constraints on the observing conditions and telluric standard stars
(e.g. their spectral type).  We note that a higher spectral  resolution also greatly improves the quality of the telluric calibration. 

To finalize the flux calibration we require good photometry, but for many GRB hosts there is a lack of high quality photometry at infrared wavelengths. This is illustrated in Section \ref{sec:060218fits}. 
The requirement for good multiband imaging also plays an important role in the acquisition process. GRB hosts are extended sources and the star forming regions of interest are blue, meaning that direct acquisition can be problematic when multiple instruments for different wavelength regions are involved, as the photocenter of the regions of interest is wavelength dependent. 
   
\item {\bf Afterglow or supernova spectra}: To maximize the scientific return it is helpful to have high resolution spectroscopy of the afterglow or SN in question. The detection of dust-tracing absorption lines (such as Na\,I, K\,I or diffuse interstellar bands) will allow comparison of the line of sight extinction properties; any detected velocity structure of absorption lines will aid in determining the position of the
GRB with respect to the star forming region (see W07 and the discussion of the UVES data above); and knowledge of the SN behaviour will help in estimating any residual supernova continuum or
nebular phase contribution to the measured fluxes (see also K{\"u}pc{\"u} Yolda\c{s}, Greiner \& Perna 2006). 

\end{itemize}

\subsection{The advantages of X-shooter \label{sec:Xshooteradv}}
The list of caveats and requirements above shows the necessity of covering a wide (restframe) wavelength range but also the inherent problems of using multiple instruments to do this.
The X-shooter spectrograph on the ESO VLT solved many of these issues. The (fixed) medium resolution of X-shooter is sufficient to deblend many of the blue Balmer lines, provide good fits
on the absorption components and to give good telluric line corrections. The very large wavelength range (in Figure \ref{fig:hhe} we show the $z = 0$ wavelength coverage of the three
X-shooter arms) means that a large number of lines are within reach, and, because of the very high efficiency of the instrument, we detect also the fainter lines. 
The architecture of the instrument is such that acquisition is done simultaneously for all three arms, meaning that the slits cover the exact same regions in the host.  The simultaneous coverage of all three arms means that some consistency in flux calibration can be achieved through the continuum as well. In cases where there is trouble bringing the relative flux calibration for each arm onto an absolute
scale, there are in some cases sufficient lines in each individual arm to allow fits for each arm separately. Using other sources (stars) in the slit for additional flux calibration, as described in Section
\ref{sec:isaac} will not be possible due to the short slit. There is still a strong requirement for accurate pre-imaging, as the X-shooter acquisition CCD only operates in the optical wavelength range.
An additional benefit of the methods shown above is that a fit on the lines in the UVB and VIS arms provides strong enough constraints on the attenuation parameters that the resulting fit can be used to 
refine flux calibration in the NIR arm, providing a way to get additional flux calibration from the science data itself.

The data above show some of these advantages already, though the exposure time for this dataset is short, and conditions were not ideal, as the data were taken in ToO mode (see Starling et al. 2010). 
Higher signal to noise observations are likely to improve the results in Section \ref{sec:xshooterfits}.

\subsection{Reddening, extinction and attenuation in GRB hosts}
Long gamma-ray bursts are accompanied by bright afterglows and in most cases by bright supernovae.  This makes GRBs ideal probes of distant galaxies: spectra can be obtained of
the afterglows to study the host ISM in absorption, and, when the afterglow has faded away, spectra of the host probe the host galaxy in emission. 
Dust properties in GRB hosts are of particular interest, because of the degeneracy of dust reddening with other parameters of interest (e.g. deviations from synchrotron models, searches for high redshift bursts through Lyman breaks etc.), the possible metallicity limits on GRB progenitors (e.g. the amount of metals locked up in dust), the influence of dust scattering on afterglow polarimetry, the 
nature of dark bursts, etc etc.

Broadly speaking the dust properties of GRB hosts can be studied using the afterglow or using the host galaxy in emission. In the first case, the amount of reddening experienced by the afterglow and
the shape of the extinction curve (or the grain size distribution)  can be derived through fits on the afterglow spectral energy distribution (e.g. Galama \& Wijers 2001; Starling et al. 2007; Schady et al. 2010 and references therein), through studies of the relative column densities of gas phase metals (e.g. Savaglio \& Fall 2004) and the wavelength dependancy of afterglow polarisation (e.g. Klose et al. 2004; Wiersema et al in prep). The distance of the gas and dust that is seen in absorption in afterglow spectra can be determined through 
variable excited metastable fine structure lines (e.g. Vreeswijk et al.~2007), excited molecular transitions (e.g. Sheffer et al. 2009), ionisation states (e.g. Prochaska, Chen \& Bloom~2006) and any signs of dust destruction by the GRB photon field (i.e. release of elements into the gas phase that were locked up in dust grains leads to observable variability of resonance lines).  

The study of dust in host galaxies can be approached by studying dust in the star forming regions hosting GRBs, which measures the influence of dust on emission line emitting hot gas (in this paper), or by SEDs of the galaxies (measuring the reddening experienced by stars).  It is clear that the $A_V$ values and extinction curves derived through all these methods are not probing the same dust columns and spatial distribution of dust, but provide complementary approaches to understand the properties of dust in star forming dwarf galaxies.

\section{Conclusions}\label{sec:conclusions}
We have presented a spectroscopic dataset of the host of GRB\,060218, which, together with already published data for a star forming tegion in the host of GRB\,100316D,
we use to get further insight into the attenuation of the emission lines by dust. Specifically we show how hydrogen and helium fluxes of many recombination transitions can be fit simultaneously
to get insight into the electron temperature and the $R_V$ of the dust. X-shooter is the ideal instrument to do this for a large sample of GRB hosts, as long as accurate calibration can be achieved. 
The resulting dust properties from recombination lines can be combined with information from absorption lines of afterglow spectra; extinction curves from afterglow spectral energy distributions
and reddening fits on integrated host galaxy spectral energy distributions to study dust formation and destruction in low metallicity dwarf galaxies.

\section*{Acknowledgments}
We thank Sandra Savaglio, Johan Fynbo, Cedric Ledoux, Palle M\o ller, Sara Ellison, Sung-Chul Yoon, Elena Pian, Rhaana Starling, Ralph Wijers, Nial Tanvir and Paul Vreeswijk for useful discussion,
and the anonymous referee for helpful comments.  
We thank the ESO staff for obtaining the data discussed here. KW acknowledges support from STFC. The financial support of the British Council and Platform Beta Techniek through the Partnership Programme in Science (PPS WS 005) is gratefully acknowledged.

\label{lastpage}

\end{document}